# $\Delta\rho$ in an SU(2)×U(1) chiral Yukawa model[*]

J.L. Alonso,[a] Ph. Boucaud[b] and A.J. van der Sijs[a]

[a]Departamento de Física Teórica, Universidad de Zaragoza,
Facultad de Ciencias, 50009 Zaragoza, Spain

[b]Laboratoire de Physique Théorique et Hautes Energies,
Université de Paris XI, 91405 Orsay Cedex, France

We report on a simulation in progress of an SU(2)×U(1) fermion-Higgs model with two *non-degenerate* fermion doublets. The Zaragoza proposal for chiral lattice fermions is used; naive fermions are studied for comparison. The aims of this simulation are a non-perturbative study of the decoupling of species doublers, of triviality bounds on the Higgs and fermion masses and of the phenomenological quantity $\Delta\rho$.

## 1. INTRODUCTION

The study of fermion-Higgs models is motivated by the attempt to define chiral gauge theories, and the Standard Model in particular, on the lattice. One assumes that corrections due to the electroweak gauge couplings can be incorporated perturbatively. The well-known obstacle to the inclusion of gauge fields is the species doubling of fermions, for which in the case of chiral theories no satisfactory solution is available [1].

In the Zaragoza proposal for chiral lattice fermions, the fermion interactions are constructed such that the species doublers are kept massless and non-interacting. The global chiral symmetry of the action is preserved (but, in the presence of gauge fields, gauge invariance is broken). This property is shared by the fermion action in the Roma I proposal, which can be shown to reduce to an action of the Zaragoza class [2].

Important issues to be investigated in chiral Yukawa models are the triviality of the Yukawa and Higgs couplings and the associated mass bounds. One would like to know in particular

[*]Contribution to Lattice 95, by A.J. van der Sijs (arjan@sol.unizar.es). Work supported by EC contracts CHRX-CT92-0051 and ERBCHBICT941067, by DGICYT project AEN 94-218 and by Acción Integrada Hispano-Francesa HF94-150B. We are indebted to F. Lesmes for collaboration at an early stage of this work.

DFTUZ/95/22 (Zaragoza), LPTHE - Orsay 95/59
September 1995

if these remain within the perturbative domain, in a model with realistic fermion content.

Here we report on a Monte Carlo study, in progress, of a fermion-Higgs model with two doublets of *non-degenerate* (Zaragoza) fermions and SU(2)×U(1) chiral symmetry. Apart from triviality issues, we study the phenomenological quantity $\Delta\rho$, which depends on the fermion mass splitting and whose leading contributions survive in the gaugeless limit of the Standard Model [3]. Its dependence on the fermion doublet multiplicity makes this quantity an interesting instrument to test the decoupling of species doublers.

## 2. $\Delta\rho$

The $\rho$ parameter is defined as the relative strength of neutral and charged current interactions at zero momentum transfer, as measured in neutrino scattering. In the Standard model, these interactions are mediated by gauge boson exchange, and at tree level one has the relation $\rho = M_W^2/(\cos^2\theta_W\, M_Z^2) = 1$.

Quantum effects cause $\rho$ to deviate from 1:

$$\Delta\rho \;=\; 1-\rho^{-1}\,. \qquad (1)$$

The dominant contribution, from ("oblique") corrections to the $W$ and $Z$ propagators, is independent of the scattering process and is called the "universal part" of $\Delta\rho$. Additional contributions come from vertex corrections and box diagrams. They are small and process-dependent. Restrict-ing ourselves to the universal part, $\rho$ can be written in terms of the propagators of the $W$ and $Z$ bosons at zero momentum,

$$\rho = \frac{1 + \Pi_W(p^2 = 0)/M_W^2}{1 + \Pi_Z(p^2 = 0)/M_Z^2}, \quad (2)$$

where $\Pi_{W,Z}(p^2)$ are the transverse parts of the vacuum polarization tensors of the $W$ and $Z$.

The leading one-loop contribution is due to the top-bottom mass splitting. It takes the form [4]

$$\Delta\rho = N_c \frac{G_F}{8\pi^2\sqrt{2}} m_t^2 = N_c \frac{1}{16\pi^2} y^2, \quad (3)$$

if the bottom mass $m_b$ is neglected compared with the top quark mass $m_t = yv$ (with $G_F^{-1} = \sqrt{2}v^2$). $N_c$ is the number of QCD colours. QCD corrections lower the right hand side by about 10 %.

This result is a manifestation of "non-decoupling" [4]: the decoupling theorem [5], which says that observable effects of heavy particles go as powers of their inverse mass, does not hold (at least, perturbatively) when the mass is generated by a Yukawa coupling.

The experimental value of $\Delta\rho$ constrains the top quark mass. The one-loop result (3) gives $m_t = \mathcal{O}(200)$ Gev, which is probably consistent with the perturbative treatment of the Yukawa coupling $y$. Alternatively, one could imagine $\Delta\rho$ to saturate as a function of $m_t$ beyond the perturbative regime, allowing a larger $m_t$. For this scenario to be viable, the maximal fermion mass (Yukawa coupling) allowed by triviality would have to lie outside the perturbative regime.

Apart from the leading result (3), there are "subleading" one-loop diagrams involving internal gauge field lines, which are relatively suppressed by powers of $g^2/y^2$ (which, by the way, is not a negligibly small number). In fact, at each loop level there is a leading term of zeroth order in $g^2$, accompanied by subleading terms which vanish in the gaugeless limit $g^2 \to 0$. Hence all the leading contributions to $\Delta\rho$ survive when the gauge fields are turned off.

We shall now focus on this gaugeless limit. This has the great benefit that calculations can be performed within the fermion-Higgs sector of the Standard Model [3]. Gauge fields are only needed to act as sources for the currents whose two-point correlation function is calculated. From the Ward identities

$$\frac{\Pi_{W^\pm,Z}(0)}{M_{W,Z}^2} \to \lim_{p^2 \to 0} \frac{\Pi_{\pi^\pm,0}(p^2)}{p^2} = \Pi'_{\pi^\pm,0}(0) \quad (4)$$

it now follows that

$$\rho = \frac{1 + \Pi'_{\pi^+}(0)}{1 + \Pi'_{\pi^0}(0)} = \frac{Z_0}{Z_+}, \quad (5)$$

where $Z_{0,+}$ are the wave function renormalization constants of the neutral and charged Nambu-Goldstone bosons (or "pions") $\pi^{0,+}$ at zero momentum [3]. We will use expression (5) to compute $\Delta\rho$ non-perturbatively in a lattice Monte Carlo simulation.

## 3. THE MODEL

Consider the SU(2) × U(1) invariant fermion-Higgs model with action

$$\begin{aligned}
S = &-\frac{k}{2} \sum_{x,\mu} \text{Tr}\left(\Phi^+(x)\Phi(x+\hat{\mu}) + h.c.\right) \\
&+ \sum_{j=1}^{2} \left[ \sum_{x,\mu} \frac{1}{2}(\overline{\psi}_j(x)\gamma_\mu \psi_j(x+\hat{\mu}) + h.c.) \right. \\
&\left. + \sum_x (\overline{\psi}_{jL}^{(1)}(x)\Phi(x)Y\psi_{jR}^{(1)}(x) + h.c.) \right]. \quad (6)
\end{aligned}$$

Here $\Phi$ is a (fixed-modulus) complex Higgs field, written as an SU(2) matrix, $\psi_{1,2}$ are two (top-bottom) fermion doublets (two because we use the Hybrid Monte Carlo algorithm) and $Y = \begin{pmatrix} y & 0 \\ 0 & 0 \end{pmatrix}$, with $y$ the top quark Yukawa coupling. The Yukawa coupling for the bottom quark(s) has been set to zero. The fermion field $\psi_j^{(1)}(x)$ in the interaction terms is defined as the average of the fields $\psi_j$ at the 16 vertices of the hypercube at $x$, ensuring the decoupling of the fermion doublers from the path integral. As a consequence of the SU(2) × U(1) symmetry of the action (6), a calculation of the leading contribution to $\Delta\rho$ in this lattice model is guaranteed to produce the correct continuum result. This was demonstrated explicitly at one loop in Ref. [6].

For comparison we will also study the model with naive fermions, which amounts to replacing $\psi^{(1)}$ by $\psi$ in the action (6). In that case

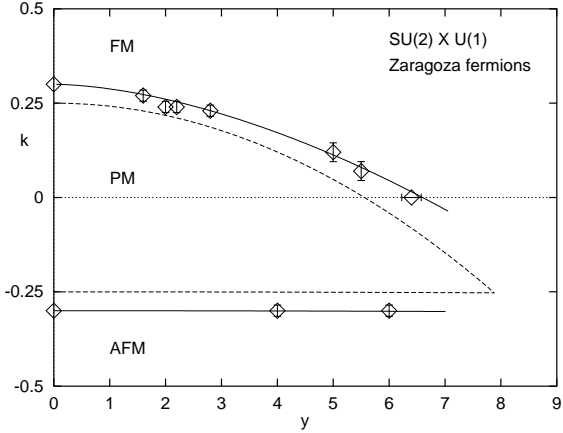

Figure 1. Weak coupling region of the phase diagram for the model of Eq. (6). Both Monte Carlo data (diamonds) and mean-field results (dashed lines) are shown.

the model describes 32 identical non-degenerate fermion doublets in the scaling region.

## 4. TOWARDS RESULTS

The relevant part of the phase diagram of the model with Zaragoza fermions is shown in Fig. 1. There are symmetric (PM) and broken (FM and AFM) phases separated by second order transitions. Monitoring of $am_t/av$ and $am_\sigma/av$ (where $v = 246$ Gev sets the scale) in the scaling region of the FM phase will provide information on the upper bounds on the top and Higgs masses imposed by triviality. By calculating $\Delta\rho$ along the way, its dependence on $m_t$ can be determined, from $m_t = 0$ up to the triviality "end point". Here we shall limit ourselves to a discussion of the determination of $\Delta\rho$ and the problems involved.

First of all, one would like to reproduce the weak-coupling result (3) numerically, in the scaling region. In Fig. 2 we compare the neutral pion propagators for naive and Zaragoza fermions. In both cases, the deviation from tree-level behaviour $1/\hat{p}^2$ is very small (recall that $Z_{\text{cont}} = 2k\, Z_{\text{latt}}$). A salient difference is observed at $\hat{p}^2 = 4, 8, 12$. As explained in Ref. [7], the "dips" in the naive case are enhancements of the propagator due to the doublers. The fact that such dips are hardly observable in the model with Zaragoza fermions is a nice illustration of the absence of doublers there, in the continuum limit.

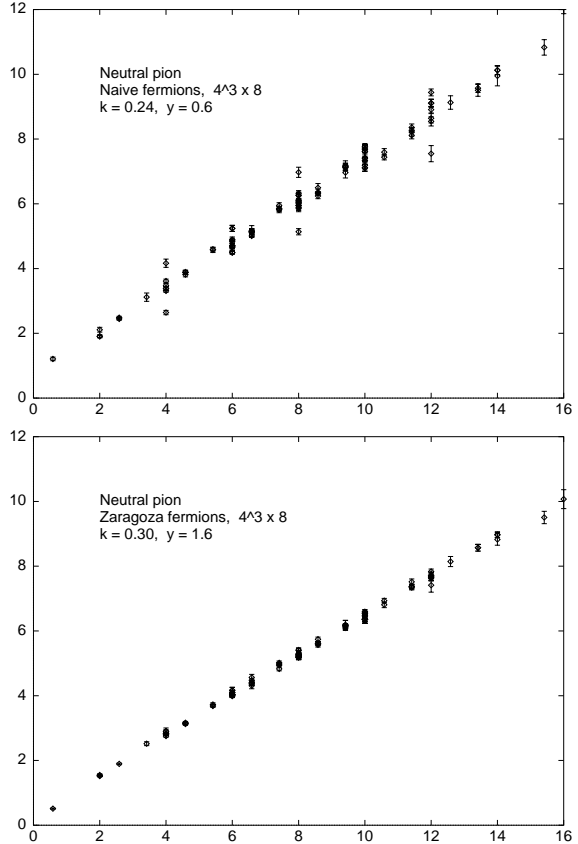

Figure 2. Monte Carlo results for the inverse lattice propagator of the neutral pion against $\hat{p}^2 = 2\sum_\mu (1 - \cos p_\mu)$, in the small-$y$ region.

The charged pion propagators, not shown here, are virtually indistinguishable from the neutral ones, except at very small momenta. Hence a very careful analysis will be required to extract $Z_0$ and $Z_+$ from the propagator data with sufficient precision to determine $\Delta\rho$ reliably.

A poor man's method to obtain numbers for the $Z$'s is to equate the data point at the smallest momentum value to $\hat{p}^2_{\min}/Z$. This is known [7] to overestimate the value of $Z$, because of the curvature of the full inverse propagator at small momenta. Let us nevertheless try this for the data in Fig. 2, to get a first impression. It is interesting to compute the quantity

$$\tilde{N} \equiv \frac{\Delta\rho}{y_R^2/16\pi^2}, \qquad (7)$$

using Eqs. (1,5), with $y_R \equiv am_t/av$. $\tilde{N}$ should come out as the multiplicity of fermion doublets, cf. Eq. (3) (32 for naive fermions, 2 for Zaragoza fermions). The poor man's method gives $224 \pm 26$ and $8 \pm 4$, respectively. These numbers are far off, which is not surprising because $\Delta\rho$ depends sensitively on the $Z$'s. It is encouraging, though, that the ratio of these numbers is large, supporting the absence of doublers in the scaling region in the Zaragoza model.

A much more accurate method, which has proven very successful in other fermion-Higgs models [7], is to fit the data with the propagator in one-fermion-loop, finite-volume renormalized perturbation theory. As for now, let us just see what can be learnt from these one-loop self-energy corrections themselves.

Fig. 3 is a plot of the ($y$-independent) quantity

$$"\tilde{N}"(p) \;=\; \frac{"\Delta\rho"(p)}{y^2/16\pi^2} \;=\; \frac{\Pi_{\pi^+}(p) - \Pi_{\pi^0}(p)}{\hat{p}^2\,(y^2/16\pi^2)} \qquad (8)$$

as a function of $\hat{p}^2$. $\Pi_{\pi^+,^0}(p)$ are the one-loop pion self-energy corrections due to *one* Zaragoza fermion doublet, calculated in a finite volume and with a finite lattice cutoff. This quantity is interesting because "$\Delta\rho$"$(0) = \Delta\rho$ and "$\Delta\rho$"$(p_{\min})$ is essentially the poor man's estimate of $\Delta\rho$.

This figure illustrates the difficulties in obtaining the correct value of $\Delta\rho$. First of all, the rapid decay (with a width $\sim (am_t)^2$) exposes the failure of the poor man's method. Second, "$\tilde{N}$"$(0)$ is much larger than the value $N = 1$ expected in the continuum limit in infinite volume. This is a finite volume effect: doubling the lattice volume reduces "$\Delta\rho$"$(p_{\min})$ by a factor 2 (even though $p_{\min}$ is smaller in the larger volume!). Third, the finite cutoff leaves its marks: in infinite volume, "$\tilde{N}$"$(0) = 0.82$ for $am_t = 0.3$, "$\tilde{N}$"$(0) = 0.51$ for $am_t = 0.8$. All these trends are similar when naive fermions are used.

Problem #1 can presumably be avoided by using the more sophisticated propagator fits men-tioned earlier. Problem #2 may be overcome by extrapolations to infinite volume. Problem #3 is probably the most difficult to handle, since relatively large values of $am_t$ are inherent to the investigation of triviality issues. On the other hand, a strong cutoff sensitivity (scaling violation) of $\Delta\rho$ might actually teach us something about the cutoff in the Standard Model or even allow us to catch a glimpse of "the other side".

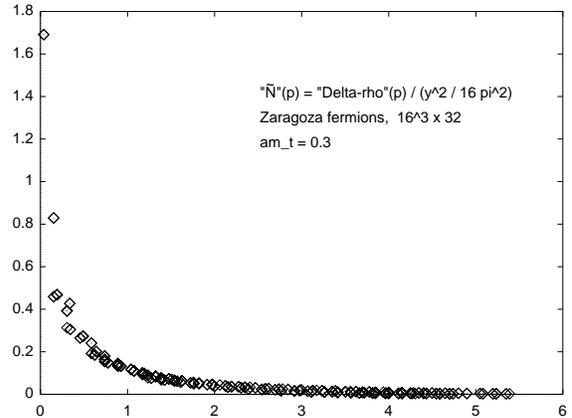

Figure 3. "$\tilde{N}$"$(p)$ against $\hat{p}^2$, cf. Eq. (8).